Polarization-Tailored Raman Frequency Conversion in Chiral Gas-Filled Hollow Core Photonic Crystal Fibers

S. Davtyan[1], D. Novoa[1], Y. Chen[1], M. H. Frosz[1], and P. St.J. Russell[1,2]
*[1]Max Planck Institute for the Science of Light, Staudtstraße 2, 91058 Erlangen, Germany*
*[2]Department of Physics, University of Erlangen-Nuremberg, 91058 Erlangen, Germany*

Broadband-tunable sources of circularly-polarized light are crucial in fields such as laser science, biomedicine and spectroscopy. Conventional sources rely on nonlinear wavelength conversion and polarization control using standard optical components, and are limited by the availability of suitably transparent crystals and glasses. Although gas-filled hollow-core photonic crystal fiber provides pressure-tunable dispersion, long well-controlled optical path-lengths, and high Raman conversion efficiency, it is unable to preserve circular polarization state, typically exhibiting weak linear birefringence. Here we report a revolutionary approach based on helically-twisted hollow-core photonic crystal fiber, which displays circular birefringence, thus robustly maintaining circular polarization state against external perturbations. This makes it possible to generate pure circularly-polarized Stokes and anti-Stokes signals by rotational Raman scattering in hydrogen. The polarization state of the frequency-shifted Raman bands can be continuously varied by tuning the gas pressure in the vicinity of the gain suppression point. The results pave the way to a new generation of compact and efficient fiber-based sources of broadband light with fully-controllable polarization state.

*Introduction.–*Optical measurements using circularly polarized light are essential in many types of non-invasive biomedical and imaging applications [1,2], as well as in tailoring the performance of magnetic memories [3] and switches [4,5]. There is also growing interest in the study of strong enantio-specific interactions of bio-molecules with circularly-polarized ultraviolet light [6,7]. Progress in these applications is hampered by a lack of versatile wavelength-tunable sources of circularly-polarized light. Potential solutions for different spectral regions include high-harmonic generation [8], the use of helical photonic metamaterials [9] or chiral plasmonic nanostructures [10]. The resulting systems, however, are in general complex, inefficient, not widely tunable and do not allow simultaneous frequency conversion, manipulation and robust transport of the polarized signals.

In this Letter we report controllable frequency conversion of circularly-polarized light in the near IR by low-threshold stimulated Raman scattering (SRS) [11] in hydrogen-filled helically-twisted hollow-core photonic crystal fiber (PCF). SRS conversion to a circularly-polarized Stokes sideband creates a coherence wave of synchronized molecular vibrations that can subsequently be used to up-shift pump light to the anti-Stokes frequency [12]. Unlike untwisted PCF, which typically exhibits weak linear birefringence, chirally twisted PCF robustly preserves circular polarization state against external perturbations. As a novel platform for chiral nonlinear photonics, twisted PCF offers excellent beam quality and broadband guidance from the vacuum ultraviolet to the mid-infrared, and opens the door to a new class of broadband light sources with controllable polarization state.

*Single-ring PCF (SR-PCF).–* The fiber used consists of a central hollow core surrounded by a ring of $N$ thin-walled capillaries, supported within a thick-walled cladding capillary (Fig. 1a,b). SR-PCF guides by anti-resonant reflection, and offers low loss guidance over a very broad spectral range [13]. 40-cm lengths of twisted and untwisted SR-PCF were used in the experiments. The core diameter in each case was $D \sim 50$ μm, the inner diameter of the capillaries $d \sim 22$ μm and their wall thickness $h \sim 820$ nm. A permanent twist of 0.5 rad/mm was introduced along the fiber axis by spinning the preform cane during fiber drawing. The $N$-fold rotationally symmetric core-surround creates a guided mode whose intensity profile is non-circular [14]. In the twisted SR-PCF this mode is forced to rotate as it propagates, introducing circular birefringence $B_C$ and offering the advantage that circular polarization states are maintained against external perturbations [15,16].

To measure the circular birefringence $B_C$ we recorded the angle of rotation of linearly-polarized light

$$\phi = \pi B_C L / \lambda = \pi (n_{RC} - n_{LC}) L / \lambda \qquad (1)$$

for a series of different fiber lengths $L$ (Fig. 1c), where $\lambda$ is the vacuum wavelength and $n_{RC}$, $n_{LC}$ are the refractive indices of RC (right-circular) and LC (left-circular) polarized modes. By fitting the data to Eq. (1), we obtained a value of $B_C \sim 3\times10^{-8}$, in good agreement with numerical modelling (Fig. 1d).

As we show in this paper, the ability to maintain circular polarization states makes gas-filled twisted SR-PCF an ideal vehicle for studying polarization-dependent SRS in a fully-controllable collinear geometry.

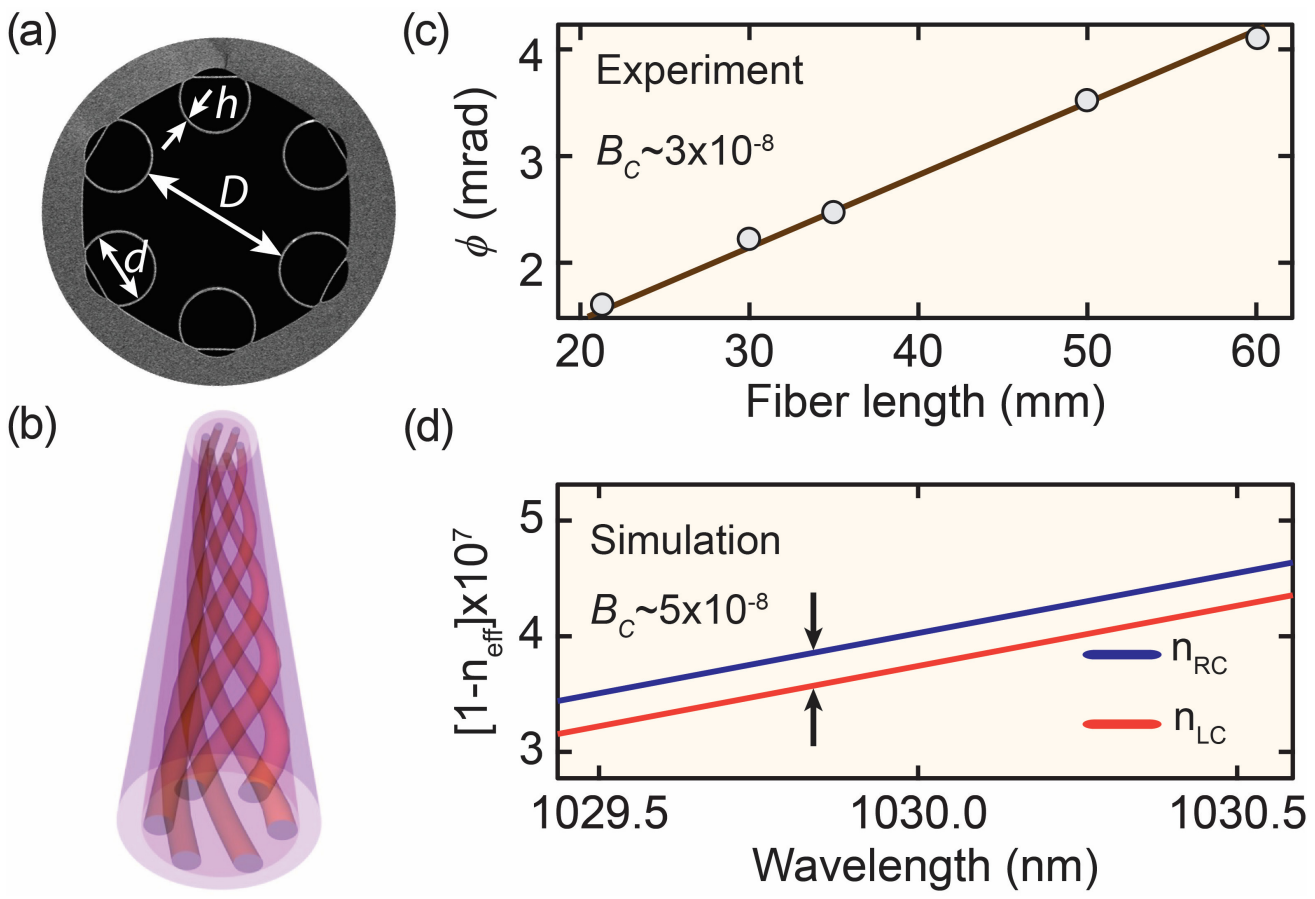

FIG. 1. (a) Scanning electron micrograph (SEM) of the transverse microstructure of the twisted SR-PCF with core diameter $D \sim 50$ µm, capillary diameter $d \sim 22$ µm, and capillary wall thickness $h \sim 0.82$ µm. The black regions are hollow and the gray regions are fused silica. The helical twist rate is 0.5 rad/mm. (b) Artist's impression of a twisted SR-PCF. Six capillaries (in red) describe a helical path around the longitudinal direction of the fiber. (c) Measurement of optical activity at 1030 nm wavelength by a series of cut-back measurements (gray discs) of the rotation angle of linear polarization. The solid line is a fit to Eq. (1), yielding $B_C \sim 3\times10^{-8}$ with a ~3% standard error. Inset: scanning electron micrograph of the microstructure of the twisted SR-PCF. (d) Numerically calculated effective indices of the fundamental RC (blue) and LC (red) modes in a twisted SR-PCF with perfect geometry.

*Theory of rotational SRS.*– Hydrogen offers high Raman gain and a relatively large rotational frequency shift ($\Omega_R/2\pi = 17.6$ THz for ortho-hydrogen at room temperature [17]). It is well-known that a circularly-polarized pump lowers the threshold intensity for rotational SRS, whereas a linearly-polarized pump favors the excitation of vibrational transitions (frequency 125 THz). The rotational Raman gain can be written $\gamma = G g_R I_P L$, where $G$ is a dimensionless coefficient that depends on the polarization state of the pump field, $g_R$ is the material gain, $I_P$ is the pump intensity and $L$ is the fiber length [18]. Assuming that the pump field is negligibly depleted during propagation, with amplitude:

$$\mathbf{e}_p = (e_{px}, e_{py}) = e_{p0}(\cos\psi,\ i\sin\psi)\,, \quad (2)$$

theory shows that the system has two gain eigenvalues [19,20]:

$$G_\pm(\psi) = [7 \pm (1 + 24\sin^2 2\psi)^{1/2}]/2\,, \quad (3)$$

whose associated Stokes field vectors are orthogonally polarized. The three-wave overlap between the coherence wave and the pump and Stokes light is higher for the + sign, yielding higher gain, so that this Stokes mode will dominate when the system is noise-seeded. For example, when $\psi = +\pi/4$ (RC pump), $G_+ = 6$ (LC Stokes) and $G_- = 1$ (RC Stokes). In contrast, when $\psi = \pi/2$ (linearly-polarized pump), $G_+ = 4$ (co-polarized Stokes) and $G_- = 3$ (orthogonally-polarized Stokes). For an elliptically-polarized pump field, the gain lies between these values. For circularly-polarized pump light, therefore, the gain is six times higher for the orthogonally polarized Stokes signal.

*Experimental set-up.*– The measurement set-up is sketched in Fig. 2a. The 1030 nm pump laser emitted linearly-polarized pulses of 1.5 ns duration at a repetition rate of 2 kHz. They were divided at a beam splitter (BS), one part being used to generate a Stokes seed signal at $\lambda_S \approx 1097$ nm (first rotational band) in a $H_2$-filled photonic band-gap fiber (PBG) with core diameter ~10 µm. A PBG-PCF was used because it offered a narrow spectral window of low loss, preventing the generation of vibrational and higher-order rotational Stokes bands (see Fig. 2b) [21].

The other part served as the pump signal ($\lambda_P$) and was delivered via a delay line to allow pulse synchronization. In contrast to noise-seeded experiments, this arrangement permits power and polarization state of pump and Stokes pulses to be controlled independently, using combinations of polarizing beam splitter (PBS) and quarter wave plate (λ/4). Both signals were combined using a dichroic mirror (DM) and launched into the fundamental mode of the $H_2$-filled twisted SR-PCF, with typical efficiencies of 80%.

The polarization state of the output field was characterized using a Stokes polarimeter [22], consisting of a removable PBS used for calibration, a quarter-wave plate placed on a motorized rotational mount, a Wollaston prism (WP) and two photodetectors (PD).

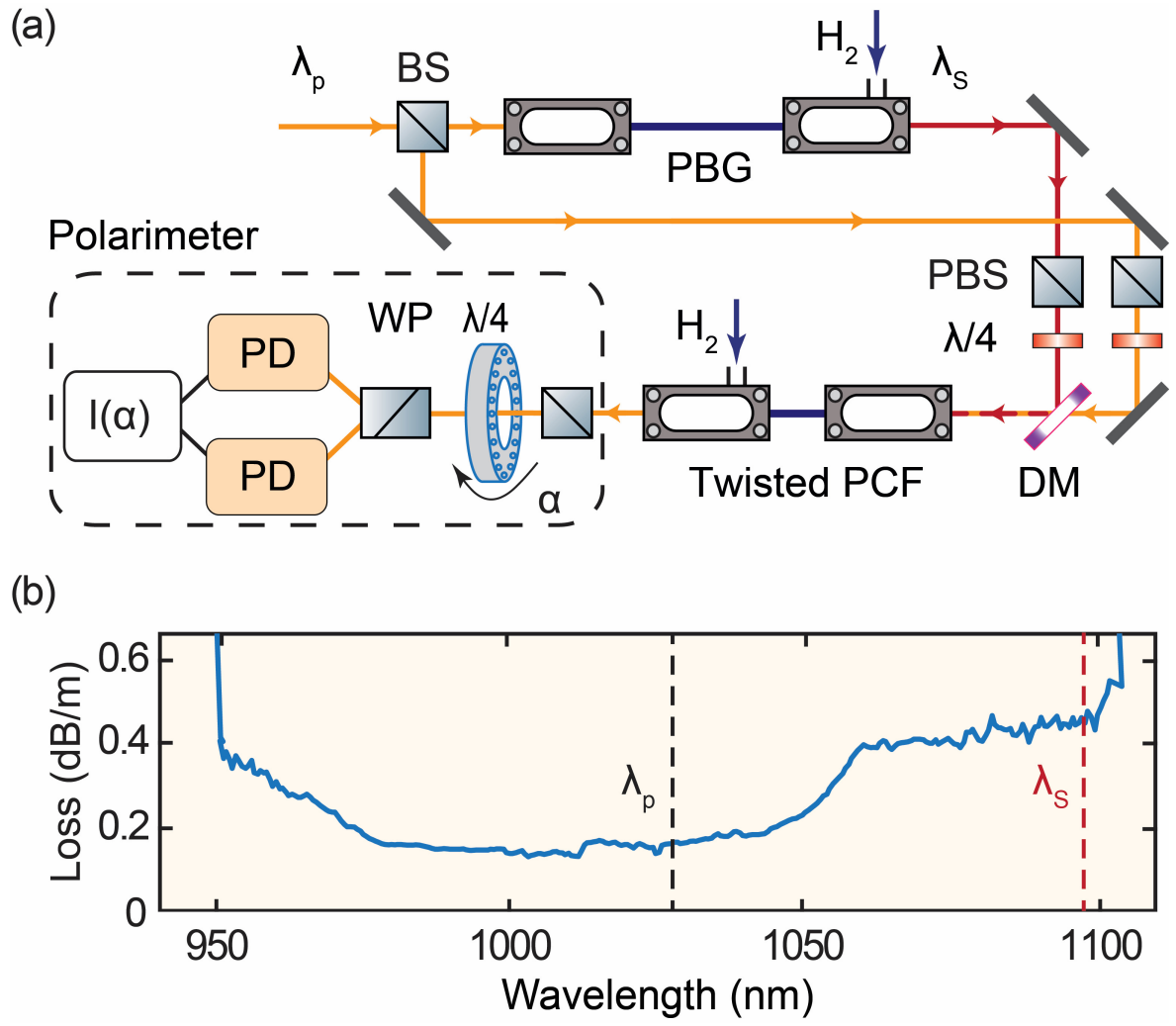

FIG. 2. (a) Scheme of the experimental set-up. (b) Loss spectrum of the photonic band-gap PCF used in the experiment.

The light intensities I(α) recorded by the photodiodes as a function of the rotation angle α allow unambiguous determination of the Stokes polarization parameters. Since our principal interest is the handedness of the polarization state, we will focus on the Stokes parameter $S_3$:

$$S_3 = (I_{RC} - I_{LC})/I_0, \quad -1 \le S_3 \le +1\,, \quad (4)$$

Where $I_{RC}$ and $I_{LC}$ are the intensities of the RC and LC components of the field and $I_0$ is the total intensity. In our measurements, $S_3$=−1,0,+1 represent the special cases of LC, linear and RC polarized light.

*Circularly-polarized fields*.– In experiments where only the pump is launched into SR-PCF (twisted or untwisted), the noise-seeded rotational Stokes signal is always found to carry spin opposite to that of the pump, in agreement with the above discussion and observed previously [23]. In untwisted SR-PCF or free space, it is practically impossible to generate a co-polarized Stokes signal, even when the system is seeded with a co-polarized Stokes signal. This is because the Raman process is extremely sensitive to small ellipticities in the launched fields, which are unavoidable in practice and result in preferential amplification of the counter-polarized Stokes.

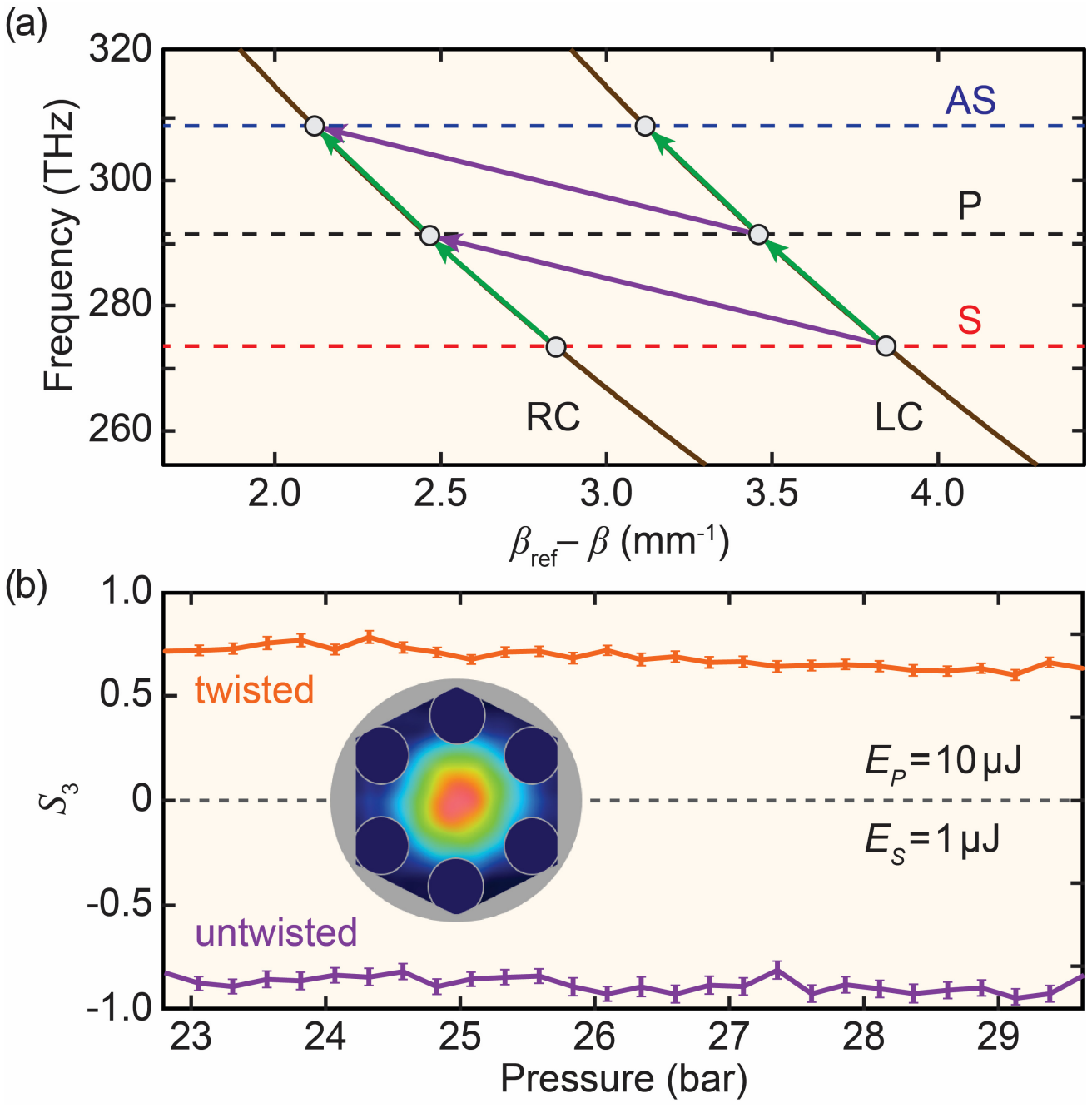

FIG. 3. (a) Dispersion diagram for the orthogonally-polarized fundamental modes of twisted SR-PCF, filled with 35 bar of $H_2$. Note that the propagation constants $\beta$ have been shifted by an amount $\beta_{ref}$ where $\beta_{ref} - \beta = 0$ at 500 THz. The green and purple arrows show the four-vectors for intra- and inter-polarization coherence waves, respectively. AS = anti-Stokes, P = pump, S = Stokes. (b) Stokes parameter $S_3$ of the generated Raman Stokes light as a function of gas pressure. $S_3$=−1,0,+1 represent LC, linear and RC polarized light. Both input pump (energy $E_P$ = 10 μJ) and Stokes (energy $E_S$ = 1 μJ) fields are RC polarized. The inset shows the measured near-field intensity profile of the Stokes signal generated in the twisted SR-PCF. The beam profile was stable during pressure scanning.

The inevitable structural nonuniformities in untwisted SR-PCF mean that guided light will develop a degree of ellipticity as it propagates, even if the input is perfectly circularly-polarized. In contrast, the circular birefringence of twisted SR-PCFs means that circular polarization state is maintained over long fiber lengths [16].

To explore the implications of this for SRS, we first launched RC co-polarized pump and Stokes fields with low energies (10 μJ pump and 1 μJ Stokes) into both twisted and untwisted SR-PCFs with similar transverse microstructures. The beat-note between these fields causes excitation of an intra-polarization coherence wave (green arrows in Fig. 3a) that in turn amplifies the Stokes seed.

Figure 3b plots the measured $S_3$ values of the transmitted Stokes signal as a function of gas pressure. In untwisted SR-PCF (solid purple line), despite the pump light being RC co-polarized, the polarization state of the output Stokes field flips to LC, mimicking the noise-seeded situation. This is because even a very small fraction of launched LC-polarized Stokes light will create inter-polarization coherence waves (purple arrows in Fig. 3a) that offer much higher gain, causing the polarization state of the transmitted Stokes signal to flip.

In twisted SR-PCF, in contrast, the polarization state of the amplified Stokes signal is very well preserved at all pressures (solid-orange line in Fig. 3b), i.e., the intra-polarization coherence waves are strong enough to ensure preservation of the spin of the generated Stokes photons, despite the gain being 6 times lower than for the competing process. To the best of our knowledge, this is the first time a circularly-polarized Stokes signal (see inset in Fig. 3(b) for a near-field transverse beam profile) has been amplified by a co-polarized collinear pump over m-long propagation distances.

In a conventional gas cell for which $B_C = 0$, the coherence wave created by a circularly-polarized pump cannot be used to generate an anti-Stokes field because otherwise spin conservation would be violated [24]. In the twisted SR-PCF, however, generation of anti-Stokes light ($\lambda_{AS} \approx 970$ nm) is influenced by the chirality. In particular, non-zero $B_C$ means that the coherence wave created by RC pump and LC Stokes pulses cannot be used to generate an LC anti-Stokes field (Fig. 3a). In practice, however, a weak RC-polarized anti-Stokes component was routinely observed in the experiments. We attribute this to small unavoidable ellipticity in the launched signals. Direct intra-polarization conversion is usually impaired by coherent Raman gain suppression [25-26], a subtle nonlinear phenomenon that can be used to precisely control the polarization of the Stokes field, as we will now discuss.

*Pressure-tunable Raman polarization*.– To observe a significant co-polarized anti-Stokes signal at the output of the twisted SR-PCF, we increased the launched pulse energies to 20 μJ (RC-polarized pump) and 2.5 μJ (RC-polarized Stokes seed). Under these conditions, a pressure scan revealed that the anti-Stokes field became more elliptical and then recovered over a narrow pressure range around 26.8 bar (Fig. 4a). At the same time the Stokes field smoothly flipped from RC to LC and then returned to RC, in a very controllable manner. In the untwisted fiber, the Stokes signal remained LC polarized, and the ellipticity of the anti-Stokes signal remained constant, at all pressures (not shown).

This intriguing effect is caused by coherent Raman gain suppression, first predicted by Shen and Bloembergen [25] and experimentally demonstrated in free-space [23] and collinear [26] geometries. In brief, the effective Raman gain in a gas can be fully suppressed if the intra-polarization coherence waves created by pump-to-Stokes conversion are identical to those annihilated in pump-to-anti-Stokes scattering (see Fig. 4b). In other words, the rates of optical phonon creation and annihilation precisely balance if the dephasing $\vartheta = \beta(\omega_{AS}) + \beta(\omega_S) - 2\beta(\omega_P)$ vanishes, where $\beta(\omega) = \omega n_{eff} / c$ is the propagation constant of the $LP_{01}$-like mode. This allows us to analytically calculate the point at which $\vartheta = 0$, provided the dispersion of the $LP_{01}$-like mode is accurately known. Although the Marcatili-Schmelzer model [27] provides a reasonable approximation (Fig. 4b, dashed curve), it was necessary to include the effect of anti-crossings between the core mode and leaky modes in the capillary walls (see the two shaded areas in Fig. 4b) [28,29]. After this was done, the calculated gain-suppression pressure agreed very well with the experimental value (Fig. 4a). In the absence of anti-crossings the model overestimated the zero-dephasing pressure by 9 bar.

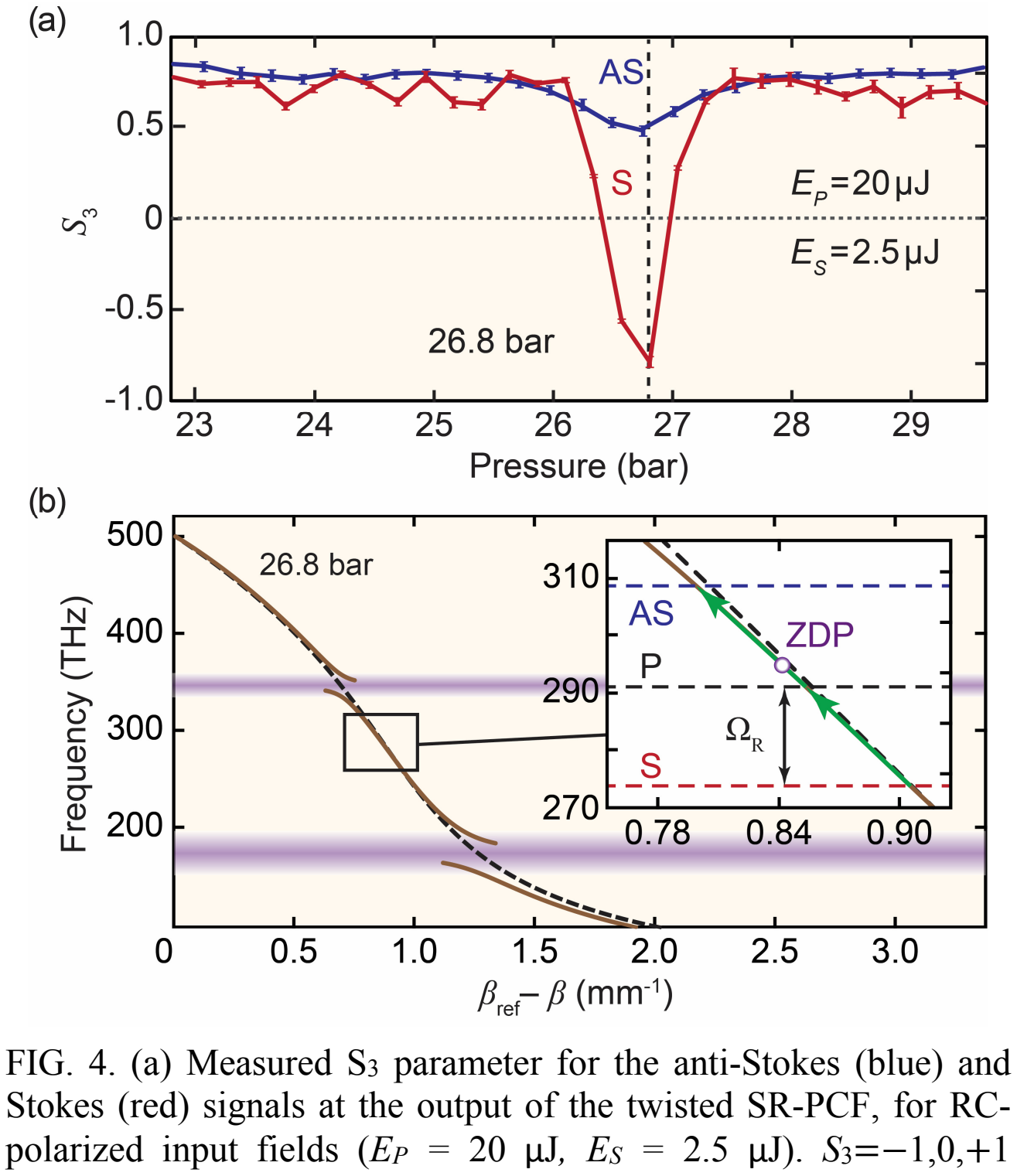

FIG. 4. (a) Measured $S_3$ parameter for the anti-Stokes (blue) and Stokes (red) signals at the output of the twisted SR-PCF, for RC-polarized input fields ($E_P$ = 20 µJ, $E_S$ = 2.5 µJ). $S_3$=−1,0,+1 represent LC, linear and RC polarized light. In the vicinity of 26.8 bar, coherent gain suppression strongly alters the output polarization states. (b) The dispersion diagram at 26.8 bar. The full-brown curve includes the effect of anti-crossings with resonances in the capillary walls (which cause loss and strong changes in dispersion) and the dashed-black curve excludes these anti-crossings. Inset: Dispersion diagram in the vicinity of the zero-dispersion point (ZDP); the coherence wave generated by pump-to-Stokes conversion is identical to that needed for pump-to-anti-Stokes conversion (green arrows).

The co-polarized pump-Stokes gain drops dramatically in the vicinity of the gain suppression point, allowing inter-polarization scattering to appear, which is not gain suppressed. This in turn permits amplification of a counter-polarized Stokes field, which significantly increases the ellipticity of the anti-Stokes signal. In addition, the polarization state of the Stokes signal flips handedness in the vicinity of the gain suppression point.

The small fluctuations in polarization state of the Stokes signal with pressure (Fig. 4a) were caused by the weakness of the Stokes seed, which was limited by the available pump laser power. Higher seed powers would provide better stability, and higher twist rates would increase $B_C$ and make the system even more resistant against external perturbations.

*Conclusions.–* The ability of twisted SR-PCF to preserve circular polarization state makes possible the selective generation of circularly-polarized Stokes and anti-Stokes light by rotational SRS in hydrogen. The polarization state of the frequency-shifted Raman bands can be continuously varied by tuning the gas pressure in the vicinity of the gain suppression point. The fibers also provide a convenient and robust means of delivering circularly-polarized light to remote locations. These characteristics, along with broad guidance from the ultraviolet to the mid-infrared [30,31] and high damage threshold, make twisted SR-PCF a promising compact platform for generation, control and delivery of broadband circularly-polarized light. In particular, with further improvements in fabrication, the approach might be extended to spectral regions such as the vacuum ultraviolet, where suitable alternatives are, to our best knowledge, not yet available.

[1] B. Kunnen, C. Macdonald, A. Doronin, S. Jacques, M. Eccles, and I. Meglinski, J. Biophotonics **8**, 317 (2015).

[2] J. F. De Boer, T. E. Milner, M. J. C. van Gemert, and J. S. Nelson, Opt. Lett. **22**, 934 (1997).

[3] A. D. Bandrauk, J. Guo, and K.-J. Yuan, J. Opt. **19**, 124016 (2017).

[4] C. D. Stanciu, F. Hansteen, A. V. Kimel, A. Kirilyuk, A. Tsukamoto, A. Itoh, and Th. Rasing, Phys. Rev. Lett. **99**, 047601 (2007).

[5] A. V. Kimel, A. Kirilyuk, P. A. Usachev, R. V. Pirasev, A. M. Balbashov, and Th. Rasing, Nature **435**, 655 (2005).

[6] K. Banerjee-Ghosh, O. Ben Dor, F. Tassinari, E. Capua, S. Yochelis, A. Capua, S.-H. Yang, S. S. P. Parkin, S. Sarkar, L. Kronik *et al.*, Science **360**, 1331 (2018).

[7] C. Meinert, I. Myrgorodska, P. de Marcellus, T. Buhse, L. Nahon, S. V. Hoffmann, L. L. S. d'Hendecourt, and U. J. Meierhenrich, Science **352**, 208 (2016).

[8] P.-C. Huang, C. Hernández-García, J.-T. Huang, P.-Y. Huang, C.-H. Lu, L. Rego, D. D. Hickstein, J. L. Ellis, A. Jaron-Becker, A. Becker *et al.*, Nat. Photonics **12**, 349 (2018).
[9] J. K. Gansel, M. Thiel, M. S. Rill, M. Decker, K. Bade, V. Saile, G. von Freymann, S. Linden, and M. Wegener, Science **325**, 1513 (2009).
[10] M. Schäferling, D. Dregely, M. Hentschel, and H. Giessen, Phys. Rev. X **2**, 031010 (2012).
[11] F. Benabid, G. Antonopoulos, J. C. Knight, and P. St.J. Russell, Phys. Rev. Lett. **95**, 213903 (2005).
[12] S. T. Bauerschmidt, D. Novoa, A. Abdolvand, and P. St.J. Russell, Optica **2**, 536 (2015).
[13] C. Wei, R. J. Weiblen, C. R. Menyuk, and J. Hu, Adv. Opt. Photon. **9**, 504 (2017).
[14] N. N. Edavalath, M. C. Günendi, R. Beravat, G. K. L. Wong, M. H. Frosz, J.-M. Ménard, and P. St.J. Russell, Opt. Lett. **42**, 2074 (2017).
[15] P. Roth, Y. Chen, M. C. Günendi, R. Beravat, N. N. Edavalath, M. H. Frosz, G. Ahmed, G. K. L. Wong, and P. S. J. Russell, Optica **5**, 1315 (2018).
[16] X. M. Xi, T. Weiss, G. K. L. Wong, F. Biancalana, S. M. Barnett, M. J. Padgett, and P. St. J. Russell, Phys. Rev. Lett. **110**, 143903 (2013).
[17] N. Bloembergen, G. Bret, P. Lallemand, A. Pine, and P. Simova, IEEE J. Quantum Electron. **3**, 197 (1967).
[18] R. Holmes and A. Flusberg, Phys. Rev. A **37**, 1588 (1988).
[19] G. V. Venkin, Yu. A. Il'inskiĭ, and G. M. Mikheev, Kvant. Electron. (Moscow) [Sov. J. Quantum Electron] **12**, 608 (1985).
[20] T. Pritchett, J. Smith, G. Mcintyre, T. R. Moore, and B. G. Oldaker, J. Mod. Opt. **46**, 949 (1999).
[21] F. Benabid, J. C. Knight, G. Antonopoulos, and P. S. J. Russell, Science **298**, 399 (2002).
[22] H. G. Berry, G. Gabrielse, and A. E. Livingston, Appl. Opt. **16**, 3200 (1977).
[23] M. D. Duncan, R. Mahon, J. Reintjes, and L. L. Tankersley, Opt. Lett. **11**, 803 (1986).
[24] R. W. Minck, E. E. Hagenlocker, and W. G. Rado, Phys. Rev. Lett. **17**, 229 (1966).
[25] Y. R. Shen and N. Bloembergen, Phys. Rev. **137**, A1787 (1965).
[26] S. T. Bauerschmidt, D. Novoa, and P. St.J. Russell, Phys. Rev. Lett. **115**, 243901 (2015).
[27] E. A. J. Marcatili and R. A. Schmeltzer, Bell Syst. Tech. J. **43**, 1783 (1964).
[28] M. Zeisberger and M. Schmidt, Sci. Rep. **7**, 11761 (2017).
[29] F. Tani, F. Köttig, D. Novoa, R. Keding, and P. St.J. Russell, Photonics Res. **6**, 84 (2018).
[30] S.-F. Gao, Y.-Y. Wang, W. Ding, and P. Wang, Opt. Lett. **43**, 1347 (2018).
[31] M. Cassataro, D. Novoa, M. C. Günendi, N. N. Edavalath, M. H. Frosz, J. C. Travers, and P. St.J. Russell, Opt. Express **25**, 7637 (2017).